\documentclass[12pt,a4paper,aps,prb,preprint]{revtex4}

\usepackage{epsfig}

\newcommand{\beq}{\begin{equation}}
\newcommand{\eeq}{\end{equation}}
\newcommand{\beqa}{\begin{eqnarray}}
\newcommand{\eeqa}{\end{eqnarray}}

\begin{document}


\thispagestyle{empty}
\vskip 3cm

\title{Study of the Damped Pendulum.} 
\author{Akhil Arora, Rahul Rawat, Sampreet Kaur \& P. Arun}
\affiliation{Department of Physics \& Electronics, S.G.T.B. Khalsa College, 
University of Delhi, Delhi - 110 007, India}
 \email{arunp92@physics.du.ac.in}

\begin{abstract}

Experiments on the oscillatory motion of a suspended bar magnet throws light
on the damping effects acting on the pendulum. The viscous drag offered by
air was found the be the main contributor for slowing the pendulum down. The
nature and magnitude of the damping effects were shown to be strongly
dependent on the amplitude. 

\end{abstract}

\maketitle


\section{Introduction}

The simple pendulum is pedagogically a very important experiment. As a
result of it's simplicity, it is introduced at high school level. The
equation of motion (EOM) of a pendulum undergoing simple harmonic motion
(SHM) is given as
\begin{eqnarray}
{d^2y \over dt^2}=-{g \over L}y=-\omega_o^2y\nonumber
\end{eqnarray}
whose solution is easily derivable and can be taught in a class which has
been introduced to calculus. The EOM can be modified to account for damping
as seen in a real pendulum and yet the equation and it's solution remains
trivial as \cite{r1}
\begin{eqnarray}
{d^2y \over dt^2} &+& \left({b \over L}\right){dy \over dt}+\omega_o^2y
=0\nonumber\\
y(t) &=& e^{-\beta t}(Acos\omega 't+Bsin\omega 't)\label{damp}
\end{eqnarray}
where ${\rm \beta=(b/2mL)}$ and ${\rm \omega'=\sqrt{\omega_o^2-\beta^2}}$.
However, this approach taken by textbooks over-simplifies the complex motion
of the pendulum and implies that only the pendulum's amplitude attenuates
with time. On the contrary along with the amplitude even the oscillation's
time period varies\cite{r2}, a feature overlooked in classroom physics and
carried forward for a long time by students. The difficulty in measuring
these variations also does not encourage routine experimentation in high
schools/ undergraduate laboratories. However, with the advent of
micro-computers such measurements can now be made easily. Most of the
experiments reported have measured the change in amplitude \cite{r2,r3,r4}
while examples of measuring variation in time period is rare\cite{r5}.
Since, both amplitude and the time period varies with successive oscillation,
one can expect the pendulum's velocity to vary with time at a given position.
While Gregory\cite{r1} used knowledge of the oscillation time period to
extract information on the pendulum's velocity, Avinash Singh {\it et 
al}\cite{r6} used a novel method to estimate the pendulum's velocity. A bar
magnet was attached to a rigid semicircular aluminum frame of radius 'L'
which pivoted about the center of the circle such that the bar magnet
oscillates through a coil kept at the mean position. As the magnet
periodically passed through the coil, it generated a series of {\it emf} pulses.
The arrangement with proper circuitry determined the peak {\it emf}.
Avinash {\it et al}\cite{r6} approximated the peak {\it emf} (${\rm
\xi_{max}}$) as
\begin{eqnarray}
\xi_{max} \approx \left({d\phi \over dt}\right)_{max}\omega_{max}\nonumber
\end{eqnarray}
where ${\rm \omega_{max}}$ is the maximum velocity as the bar magnet passed
through the mean position. This method has it's advantage when one proposes
to study the damping effects in a pendulum. Most of the works studying the
variation in oscillation amplitude\cite{r4,r10,r11} with time have the pendulum's 
suspension connected to a variable resistance (potentiometer) which
introduces a sliding friction in the pendulum's motion. Complex mathematics
with assumption that all damping contributors act independently is then used
to filter out information of each contribution. Wang et al\cite{r12} used a
novel but costly method using Doppler effect to monitor the position of the
pendulum to study it's damping. Thus, Avinash et al\cite{r6} provides a
interesting yet cheap method to study the damped pendulum. While they rightly 
pointed out that several
parameters of the experiment such as velocity and strength of the magnet and
the number of turns in the coil can be varied, they did not
explicitly discuss them theoretically or study these factors experimentally.
Hence, in this manuscript, we have furthered the study made in ref 6 and have
tried to address these issues.

\section{Experimental Setup}

Our pendula was made by suspending a standard bar magnet by a cotton thread. The
tread was fastened to a small hook drilled into one pole of the bar magnet.
The length of the bar magnet ({\it 2l}) was 7cm and the cotton
thread (${\rm L_s}$) used was 53cm long. A coil of 1000 turns was kept near the
pendula's mean position at a distance 'd' from the magnet's lower pole (see
fig 1). The magnetic field at point 'A' is evaluated by 
\begin{eqnarray}
B={\mu_o m \over 4\pi}\left[{1 \over BA^2}-{1 \over CA^2}\right]\label{eq1}
\end{eqnarray}
where m is the dipole moment. 'AC' and 'BA' can be written in terms of the pendulum's
position (angle ${\rm \Theta}$) using the cosine law. That is,
\begin{eqnarray}
BA^2 &=& OB^2+OA^2-OB.OAcos\Theta\nonumber\\
AC^2 &=& OC^2+OA^2-OC.OAcos\Theta\nonumber
\end{eqnarray}
\begin{figure}[h]
\begin{center}
\epsfig{file=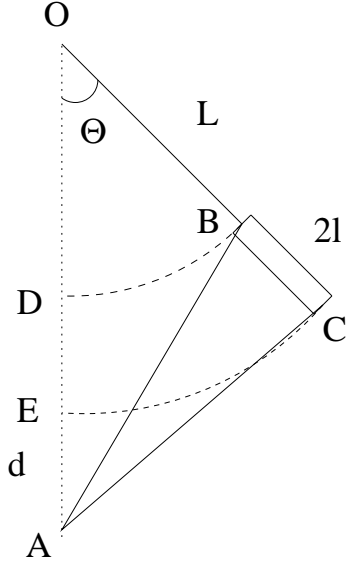,width=2in}
\vskip -.3cm
\caption{Pendulum with the mass being replaced by a bar magnet. The 
detecting coil is at 'A'.}
\vskip -1cm
\end{center}
\label{fig:1}
\end{figure}
where
\begin{eqnarray}
OC &=& L_s+2l\nonumber\\
OB &=& L_s \nonumber\\
OA &=& L_s+2l+d\nonumber
\end{eqnarray}
hence,
\begin{eqnarray}
BA^2 &=& L_s^2+(L_s+2l+d)^2-L_s.(L_s+2l+d)cos\Theta\nonumber\\
AC^2 &=& (L_s+2l)^2+(L_s+2l+d)^2-(L_s+2l).(L_s+2l+d)cos\Theta\nonumber
\end{eqnarray}

Based on the assumption that {\it 2l} and {\it d} are relatively small
compared to {\it ${\rm L_s}$}, the higher powers of {\it 2l} and {\it d} can be
neglected. Hence, eqn(\ref{eq1}) can be written as
\begin{eqnarray}
B &\approx& {\mu_o M \over 4\pi}\left[{2l \over
L_s^2(L_s+4l+d)(2-cos\Theta)}\right]\nonumber
\end{eqnarray}
The induced {\it emf} is proportional to the rate of change in the number of
magnetic lines cutting the coil. 
\begin{eqnarray}
{dB \over dt} \approx -{\mu_o M \over 4\pi}\left[{2l \over
L_s^2(L_s+4l+d)}\right]\left[{sin\Theta \over (2-cos\Theta)^2}\right]{d \Theta
\over dt}\nonumber
\end{eqnarray}
Based on this, the respective induced {\it emf} can be written as 
\begin{eqnarray}
\xi=-N{dB \over dt} \approx {\mu_o MN \over 4\pi}\left[{2l \over
L_s^2(L_s+4l+d)}\right]\left[{sin\Theta \over (2-cos\Theta)^2}\right]{d \Theta
\over dt}\label{eq8}
\end{eqnarray}
where {\it N} is the number of turnings in the coil. 
Eqn(\ref{eq8}) can be written in a compact form 
\begin{eqnarray}
\xi = \xi_o\left[{sin\Theta \over
(2-cos\Theta)^2}\right]\left({d\Theta \over dt}\right)\label{eq10}
\end{eqnarray}
where
\begin{eqnarray}
\xi_o= {\mu_o MN \over 4\pi}\left[{2l \over L_s^2(L_s+4l+d)}\right]\label{eq8a}
\end{eqnarray}

\begin{figure}[t]
\begin{center}
\epsfig{file=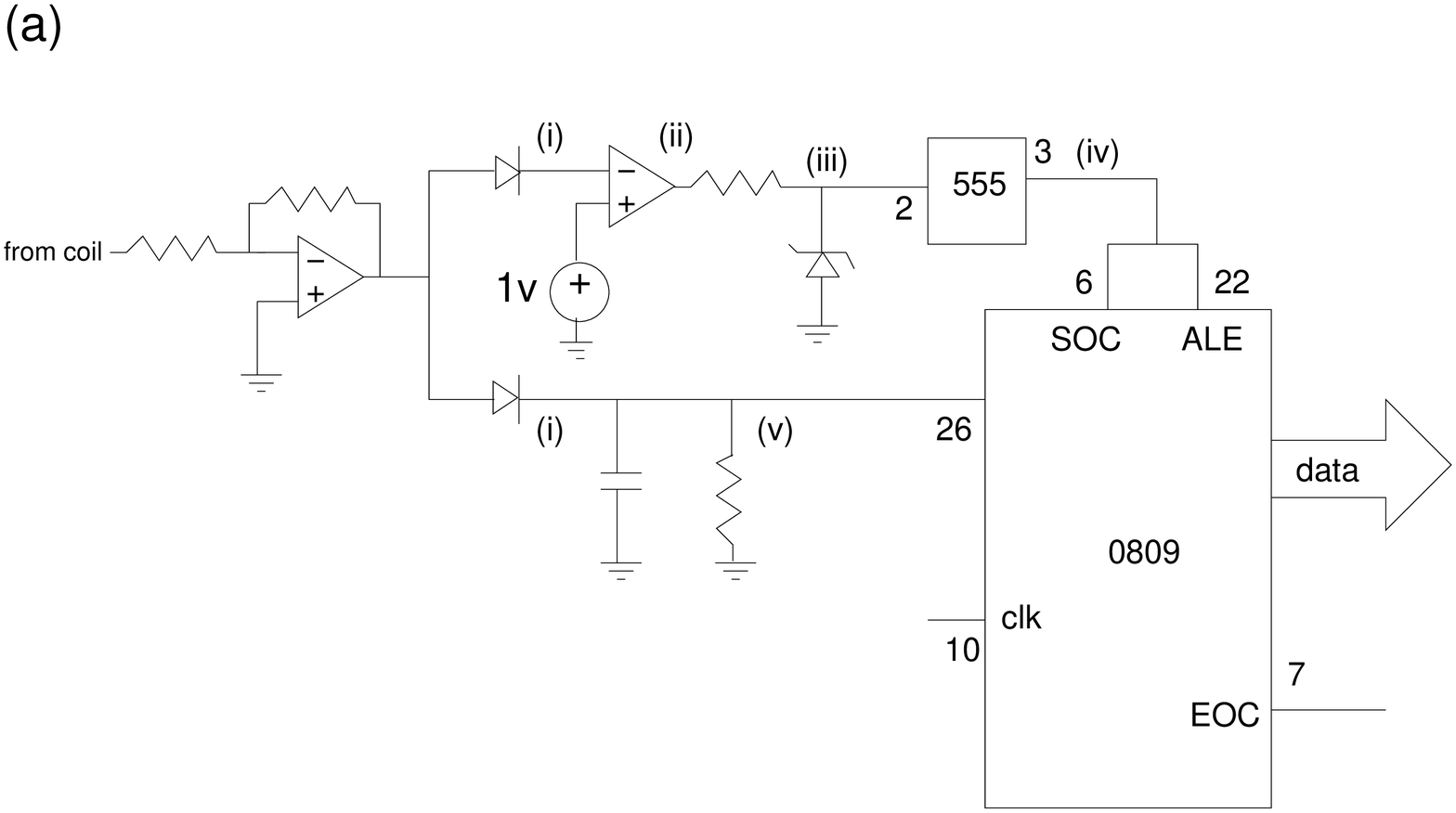,width=6in}
\epsfig{file=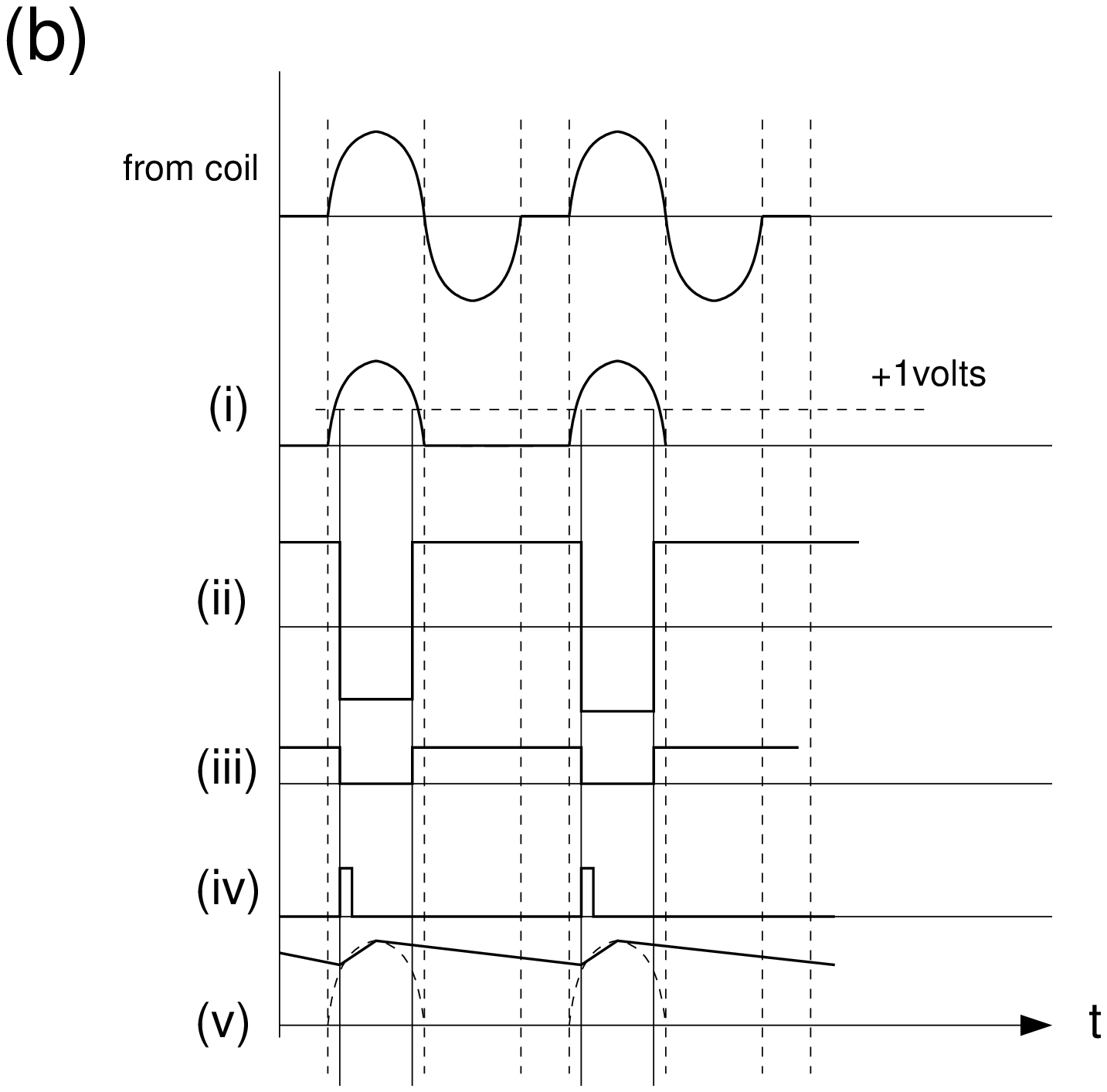,width=2.75in}
\vskip -.3cm
\caption{The (a) schematic diagram of the circuit used and the (b) important 
waveforms at the points marked in the circuit.}
\vskip -1cm
\end{center}
\label{fig:2aa}
\end{figure}

Thus, as the distance between the magnet and detecting coil is increased,
the induced {\it emf} decreases. Infact the induced {\it emf} is quite weak
and is amplified by an op amp circuit. The high input impedance of the IC741
opamp ensures that a true measurement of the {\it emf} is made. To
digitalise this analog signal (see fig 1, ref 6) using an Analog to Digital 
Convertor ADC-0809 (see fig 2a), the
amplified output is rectified and the peak value is held by charging a
capacitor. The capacitor is discharged via a large resistance so that it
retains the peak value till the next peak value arrives.

\begin{figure}[h]
\begin{center}
\epsfig{file=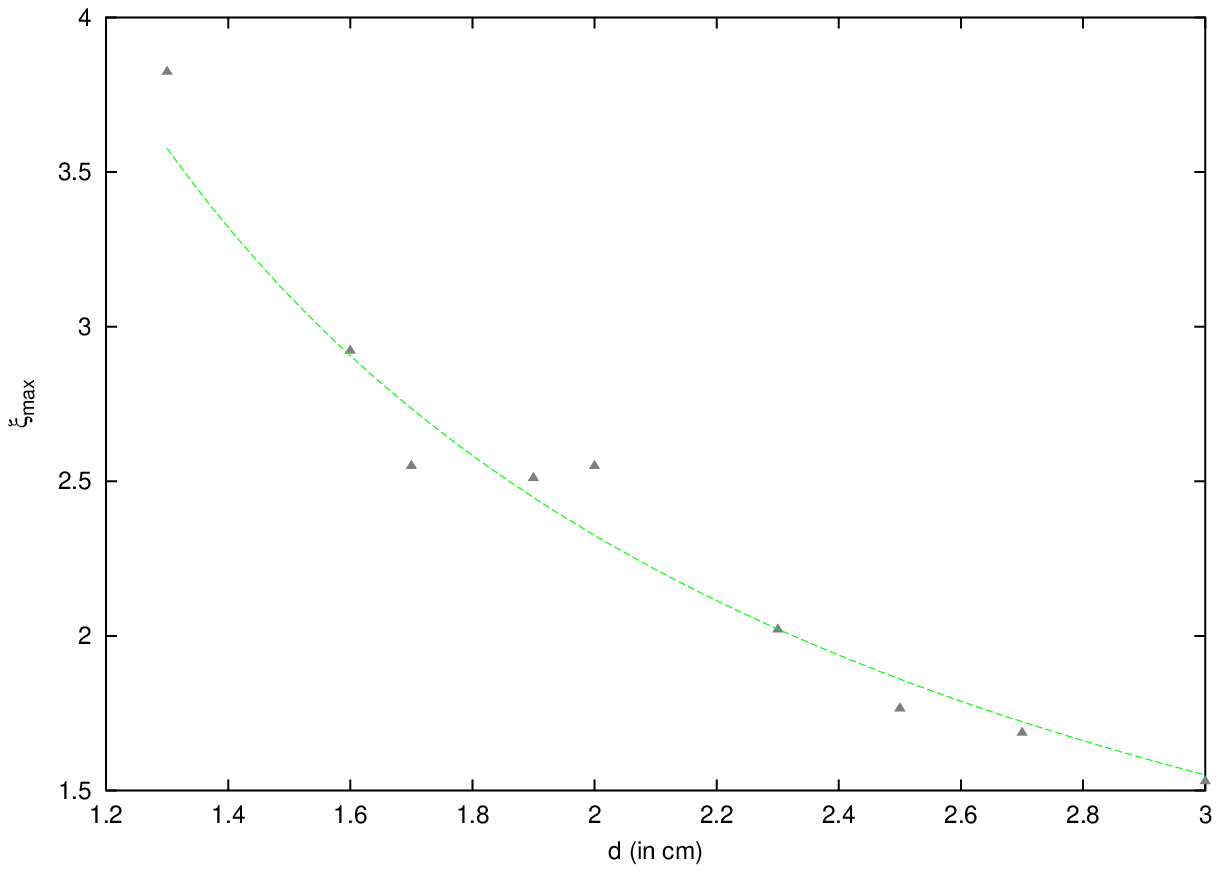,width=4in}
\vskip -.3cm
\caption{Variation in the maximum induced {\it emf} with increasing distance
between the coil and the magnet.}
\vskip -1cm
\end{center}
\label{fig:33}
\end{figure}

We require the ADC to start conversion once
the peak value is attained by the capacitor. This implies a synchronization
between the input {\it emf} pulses and the ADC's start of conversion (SOC)
pulses. To achieve this synchronization it is best to generate the required
SOC pulse by wave-shaping the input itself. The amplified input after
rectification is fed to a comparator which compares to +1v. This is to avoid
spurious/accidental triggering due to noise. The infinite gain results in
pulses with sharp edges. The width of these pulses are approximately ${\rm
T_o/4}$ (for our pendula ${\rm \approx 390ms}$). This would be too large for
serving as a SOC and hence is reduced to a ${\rm 5\mu s}$ pulse using a
monostable timer made with IC555\cite{r8}. The sequencing and
synchronization can be understood from the various waveforms shown in fig
2b. The designed circuit digitalises the analog {\it emf} and on
completion sends an EOC to the computer or microprocessor kit (in case of a
microprocessor this is done through a programmable I/O IC8155 chip, details
of which can be found in the book by Goankar\cite{r7}) which then reads the
eight bit data and stores it for retrivial. This project was done using an
8085 microprocessor kit. The programme and flowchart used is detailed in the
Appendix.

The reliability of our circuit can be tested by measuring the maximum {\it
emf} induced in the coil for varying distances 'd'. Eq(\ref{eq10}) shows
that the measured maximum {\it emf} would be directly proportional to ${\rm
\xi_o}$ which inturn is inversely proportional to 'd' (see eqn \ref{eq8a}).
Fig 3 shows the variation in the experimentally determined ${\rm \xi_{max}}$
with 'd'. While the inverse nature is evident, the value of ${(L_s+4l)}$ as
returned by curve fitting eqn(\ref{eq8a}) on our data is substantially off
mark from the actual lengths. This is expected since eqn(\ref{eq10}) and
(\ref{eq8a}) are very simplified approximations.

\section{Variation of Induced Emf with Initial Displacement}
\subsection{While undergoing undamped oscillation}

The velocity of an undamped pendulum undergoing SHM
is given as
\begin{eqnarray}
\left({d\Theta \over dt}\right)=\omega_o\sqrt{\Theta_m^2-\Theta^2}\nonumber
\end{eqnarray}
where ${\rm \omega_o=\sqrt{g/(L_s+2l)}}$ is the frequency of oscillation and
${\rm \Theta_m}$ is the initial displacement given to the pendulum.
Therefore, the emf induced by our pendulum undergoing undamped SHM would be 
given as (using eq \ref{eq10})
\begin{eqnarray}
\xi \approx \omega_o \xi_o\left[{sin\Theta \sqrt{\Theta_m^2-\Theta^2}\over
(2-cos\Theta)^2}\right]\label{eq7}
\end{eqnarray}
The variation in induced {\it emf} with time of an 
undamped pendulum undergoing SHM is as calculated using eqn(\ref{eq7}) is 
shown in fig(4). The maximum angular displacement used to generate the graph 
using eqn(\ref{eq7}) was ${\rm 5^o}$. The emf pulse shown in fig 4 is only
for half a cycle starting from one extreme position to the opposite extreme.
As the magnet approaches the coil, the flux increases and as it crosses the
mean position, the emf is negative since the magnet is receding from the
coil. Eventhough the velocity (${\rm d\Theta/dt}$) is maximum at the mean
position, since the variation in flux (${\rm d\phi /dt}$) is zero, the
induced emf is zero as the pendulum passes the mean position.

\begin{figure}[h]
\begin{center}
\epsfig{file=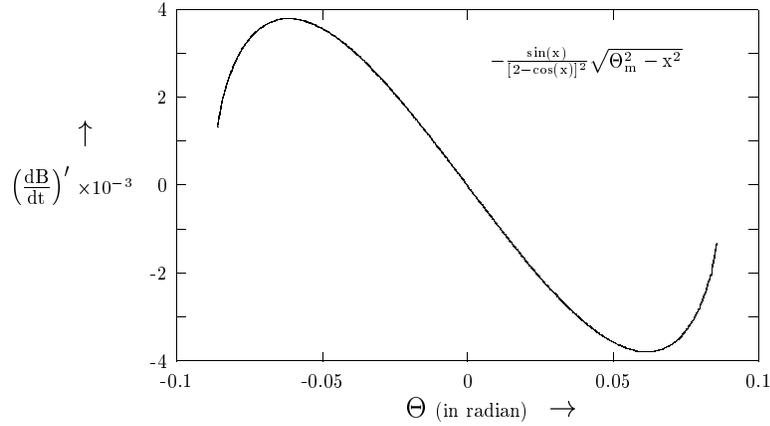,width=4in}
\vskip -.3cm
\caption{A measure of the induced {\it emf} with
oscillating angle. The graph was generated using eqn(\ref{eq7}) with ${\rm
\Theta_m=5^o}$.}
\vskip -1cm
\end{center}
\label{fig:2}
\end{figure}
The position of the pendulum when the maximum emf is generated (between 
${\rm 0< \Theta < \Theta_m}$) can be found as a problem of maxima and minima
\begin{eqnarray}
{d\xi \over d\Theta} &=& \xi_o\left\{{(2-cos \Theta)^2\left[sin\Theta {d \over
d\Theta}\left({d\Theta \over dt}\right)+cos\Theta 
\left({d\Theta \over dt}\right)\right] -(2-cos\Theta)sin^2\Theta \left({d\Theta
\over dt}\right)\over
(2-cos\Theta)^3}\right\}\nonumber\\
&=& \xi_o\left\{{(2-cos \Theta)\left[sin\Theta {d \over
d\Theta}\left({d\Theta \over dt}\right)+cos\Theta 
\left({d\Theta \over dt}\right)\right] -sin^2\Theta \left({d\Theta
\over dt}\right)\over
(2-cos\Theta)^2}\right\}\nonumber
\end{eqnarray}
For cases of small angle oscillations eqn(\ref{eq11}) reduces to
\begin{eqnarray}
{d\xi \over d\Theta} &=& \xi_o\left[\Theta {d \over
d\Theta}\left({d\Theta \over dt}\right)+(1-\Theta^2) \left({d\Theta
\over dt}\right)\right]\label{eq11a}
\end{eqnarray}
\begin{eqnarray}
{d\xi \over d\Theta} = \omega_o \xi_o\left[{-2\Theta^2 \over
2\sqrt{\Theta_m^2-\Theta^2}}+
(1 -\Theta^2) \sqrt{\Theta_m^2-\Theta^2}\right]\nonumber
\end{eqnarray}
\begin{eqnarray}
\Theta^2 = (1 -\Theta^2)(\Theta_m^2-\Theta^2)\nonumber
\end{eqnarray}
Solving the quadratic equation
\begin{eqnarray}
\Theta^4-(2+\Theta_m^2)\Theta^2+\Theta_m^2=0\nonumber
\end{eqnarray}
we have
\begin{eqnarray}
\Theta_{peak}=\pm{\Theta_m \over \sqrt{2}}\label{eq18}
\end{eqnarray}
Since eqn(\ref{eq11a}) was used to determine position of extrema, the above
condition is only valid for undamped small angle oscillations.
The maxima as per this condition for magnet oscillating through ${\rm
\Theta_m=5^o}$ occurs at ${\rm \pm 0.0617}$ radians (or ${\rm \pm 3.53^o}$).
The maximum {\it emf} that is induced, hence is (use eqn \ref{eq7})
\begin{eqnarray}
\xi_{max}=\omega_o \xi_o {\Theta_m \over \sqrt{2}}\times {sin {\Theta_m \over
\sqrt{2}} \over \left(2-cos{\Theta_m \over \sqrt{2}}\right)^2}\nonumber
\end{eqnarray}
Since, these equations and conditions are essentially valid for small
angles,
\begin{eqnarray}
\xi_{max}=\left({\omega_o \xi_o \over 2}\right)\Theta_m^2 \label{eqn25new}
\end{eqnarray}
However, a physical pendulum is prone to damping and hence in the next
section we investigate as to how the maximum induced emf varies with initial
displacement for a damped pendulum. 

\begin{figure}[h]
\begin{center}
\epsfig{file=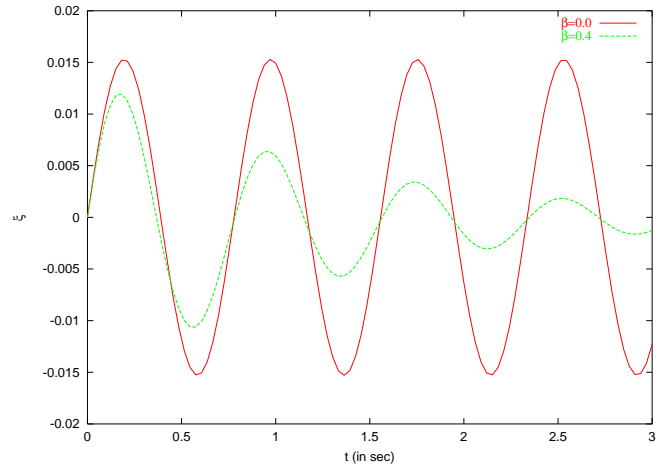,width=3.5in}
\vskip -.3cm
\caption{Variation of induced emf when oscillation is damped (${\rm
\beta=0.45s^{-1}}$) is compared with the case of no damping (i.e. ${\rm
\beta=0.0s^{-1}}$).}
\vskip -1cm
\end{center}
\label{fig:3}
\end{figure}

\subsection{While undergoing damped oscillation}

\begin{figure}[h]
\begin{center}
\epsfig{file=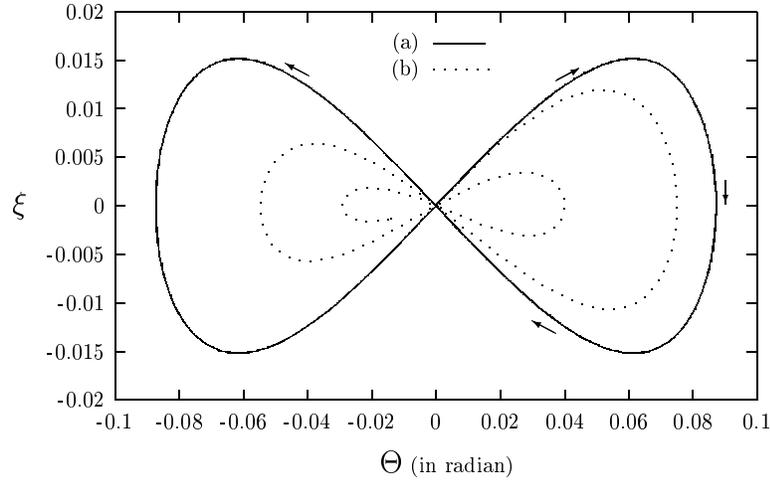,width=4in}
\vskip -.3cm
\caption{The variation of induced emf of fig 3 is plotted w.r.t. angular
position of the pendulum (i.e. ${\rm \Theta}$) for the cases (a) ${\rm
\beta=0.0s^{-1}}$ and (b) ${\rm \beta=0.45s^{-1}}$.}
\vskip -1cm
\end{center}
\label{fig:4}
\end{figure}

We have already stated in our introduction that the damped motion described
by eqn(\ref{damp}) exhibits how the pendulum's oscillation amplitude decreases
exponentially with time. The EOM whose solution is given by eqn(\ref{damp}) 
describes a linear system. The solution can be further trivialized without
losing any generality as
\begin{eqnarray}
\Theta=\Theta_me^{-\beta t}sin(\omega' t)\label{eq20}
\end{eqnarray}
from which the velocity can be calculated as
\begin{eqnarray} 
{d \Theta \over dt} &=& \omega' \Theta_me^{-\beta t}cos(\omega' t)-
\beta \Theta_me^{-\beta t}sin(\omega' t)\nonumber\\
&=& \Theta_me^{-\beta t}[\omega' cos(\omega' t)- \beta sin(\omega'
t)]\label{eq22}
\end{eqnarray}

Substituting the above expression in eqn(\ref{eq10}) we obtain the relation
showing the variation of induced {\it emf} with time. This variation is
shown in fig 5. It is also clear from the figure that the peaks in the
induced {\it emf} occurs at ${\rm \omega t=(2n+1){\pi \over 4}}$. Hence, the 
angles at which maxima occur in general is written as
\begin{eqnarray}
\Theta_{peak}=\pm {\Theta_m \over \sqrt{2}}e^{-{(2n+1)\pi \over 4tan
\phi}}\label{eq23}
\end{eqnarray}
where ${\rm tan \phi =\omega' /\beta}$.
Our circuit is designed only to measure peak {\it emfs} at
n=0,2,4,6....., where only the positive solutions of eqn(\ref{eq23}) would
contribute. Using our condition on eqn(\ref{eq22}) and eqn(\ref{eq10}) we have
\begin{eqnarray}
\left({d\Theta \over dt}\right)_{peak}=(\omega'-\beta){\Theta_m \over
\sqrt{2}}e^{-{(2n+1)\pi \over 4tan \phi}}\nonumber\\
\xi_{peak}=(\omega'-\beta)\xi_o\left[{sin 
{\Theta_m \over \sqrt{2}}e^{-{(2n+1)\pi \over 4tan\phi}}
\over (2-cos {\Theta_m \over \sqrt{2}}e^{-{(2n+1)\pi \over 4tan\phi}})^2 }\right]
{\Theta_m \over \sqrt{2}}e^{-{(2n+1)\pi \over 4tan \phi}}\label{eq24}
\end{eqnarray}
For small angle oscillations eqn(\ref{eq24}) reduces to
\begin{eqnarray}
\xi_{peak}={(\omega'-\beta)\xi_o \over 2}\Theta_m^2 
e^{-{(2n+1)\pi \over 2tan \phi}}\label{eq25}
\end{eqnarray}

\begin{figure}[h]
\begin{center}
\epsfig{file=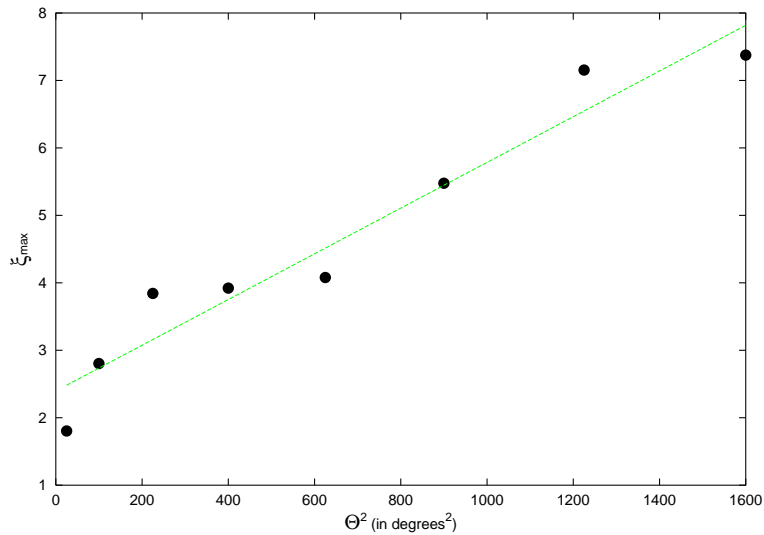,width=4in}
\vskip -.3cm
\caption{The variation of induced emf with the initial angular displacement
of the pendulum. It shows the expected parabolic dependence 
(i.e. ${\rm \Theta^2_m}$). }
\vskip -1cm
\end{center}
\label{fig:5aa}
\end{figure}

The variation in {\it emf} (seen w.r.t time in fig 5) when viewed w.r.t
oscillating angle ${\rm \Theta}$ shows how the peak
position decreases (eq \ref{eq23}) as also the amplitude of the maximum induced
{\it emf} decreases (eqn \ref{eq25}) with each half cycle.
Eqn(\ref{eqn25new}) and eqn(\ref{eq25}) shows that the maximum {\it emf}
induced for damped pendulums undergoing SHM is directly proportional to the 
square of maximum angular
displacement given to the pendulum. We have recorded the first maxima
reading (i.e. n=0) of the induced {\it emf} for various angles upto ${\rm
40^o}$. The linear relation between ${\rm \xi_{max}}$ and ${\rm \Theta_m^2}$
is evident. Before commenting further, it must be recollected that eq(1) is
valid for small angle oscillations, i.e. for ${\rm \Theta_m < 5^o}$, yet a
good linearity is obtained till ${\rm \Theta=40^o}$. 

Experimental data for ${\rm \Theta_m \geq 45^o}$ deviate markedly from this
linear trend. Remember eqn(\ref{eq23}) was obtained with the assumption that
the pendula's motion is described by eqn(\ref{eq20}). This equation describes
the motion of a pendulum oscillating in a viscous medium with small
velocity. It would be shown in the next section that the pendulum's velocity
is quite appreciable for ${\rm \Theta_m \geq 45^o}$ and hence it's motion is
not described as in eqn(\ref{eq20}), explaining the departure for linearity. 

\section{Results and Observations}

\begin{figure}[h]
\begin{center}
\epsfig{file=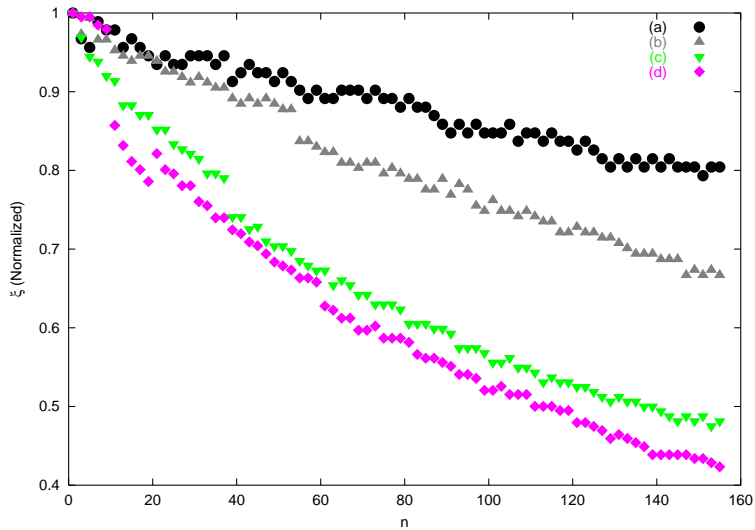,width=4in}
\vskip -.3cm
\caption{The variation in peak induced emf measured with each oscillation is 
shown for initial displacements (${\rm \Theta_m}$) (a) $\rm 5^o$, (b) $\rm 30^o$, 
(c) $\rm 55^o$ and (d) $\rm 65^o$.}
\vskip -1cm
\end{center}
\label{fig:5}
\end{figure}

Our prelimary measurements are in good correspondence with commonly known notions
and hence we proceed to investigate further the nature of damping in our
pendula. It should be noted that the amount of damping and it's nature are
strongly pendula dependent and all results reported here are specific to our
experiment and can not be taken as general. We have recorded the maxima in 
induced {\it emfs} for 80 oscillations for various initial displacements.
Since for each oscillations, we
get two positive maxima in induced {\it emf}, figure 8 shows the variation
in maxima reading of induced {\it emf} for 160 peaks.

\begin{figure}[h]
\begin{center}
\epsfig{file=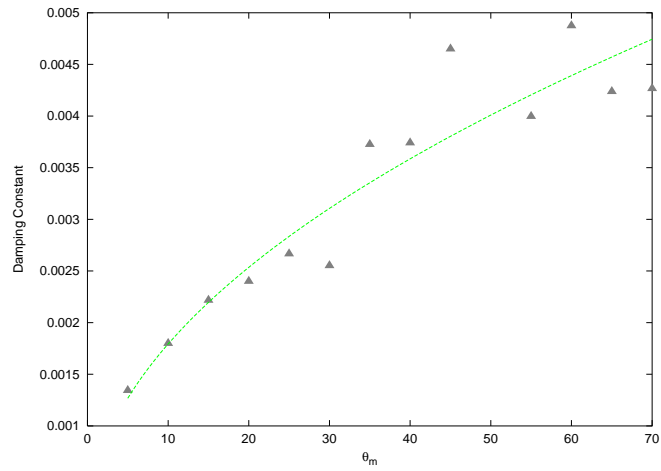,width=3.5in}
\vskip -.3cm
\caption{The variation decay constant (b of general equation ${\rm
ae^{-bn}}$) of the exponentially decaying region. The continuous line is the 
parabolic fit for the data points (${\rm 0.00056\sqrt{\Theta_m}}$).}
\vskip -1cm
\end{center}
\label{fig:5ya}
\end{figure}

\begin{figure}[h]
\begin{center}
\epsfig{file=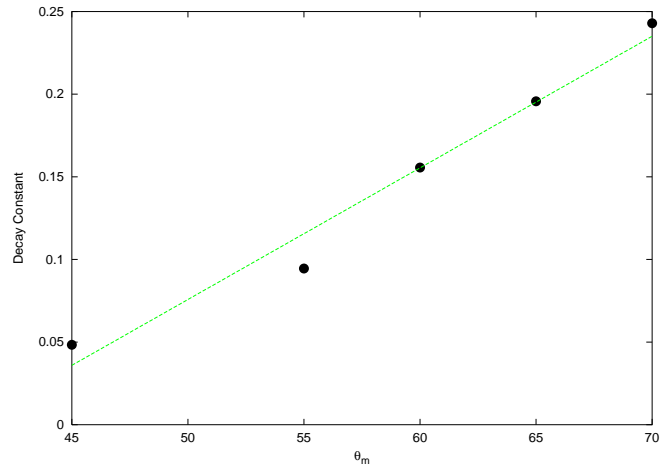,width=3.5in}
\vskip -.3cm
\caption{The variation decay in 'b', a measure of decay in the early 
oscillations when pendulum was set in motion with displacements ${\rm >45^o}$.}
\vskip -1cm
\end{center}
\label{fig:5xa}
\end{figure}

A general expression ${\rm ae^{-bn}}$ was fitted to the data of ${\rm
\xi_{max}}$
w.r.t n using a standard and freely available curve fitting software called
"Curxpt v3.0". Good fits were obtained for oscillations set by initial
displacements upto ${\rm 40^o}$. The exponential fall in {\it emf} is
indicative of the rate of loss of energy from the oscillating system. It
indicates the loss to follow the relation
\begin{eqnarray}
{dE \over dt} \propto -E\nonumber
\end{eqnarray}
or
\begin{eqnarray}
{dE \over dt} = -bE\nonumber
\end{eqnarray}
This indicates that the velocity is low and hence the damping/resistive force
acting on the pendulum is proportional to the velocity\cite{r9}. Figure 9
shows the variation of the decay constant 'b' (of ${\rm ae^{-bn}}$) with
respect to the maximum displacement (${\rm \Theta_m}$) given to the
pendulum. The graph indicates that as ${\rm \Theta_m}$ increases, the
velocity with which the pendulum moves increases with which the damping
constant increases.

For displacement angles beyond ${\rm 45^o}$ (${\rm \Theta_m \geq 45^o}$), 
eventhough visual examination of the curves in fig 8 suggests an exponential 
damping, the data points do not fit
to an exponential fall relation. A detailed examination suggests a more complex 
process is taking place with initial damping being sharper. Infact the
initial 25-30 data points fit to ${\rm a/n^b}$. The data beyond this fit to
the exponential fall equation. The ${\rm a/n^b}$ fit corresponds to the
damping force being proportional to higher powers of velocity (${\rm v^\gamma}$,
where ${\rm \gamma >1}$) and in turn the rate of energy loss also being 
proportional to higher terms of energy.
That is, the rate of energy loss for our pendulum set into oscillations with a
displacement angle ${\rm >45^o}$ is given as
\begin{eqnarray}
{dE \over dt} = -\alpha E^{1+b \over b}\nonumber
\end{eqnarray}
Figure 10 plots 'b' versus ${\rm \Theta_m}$. The power term (b) is being
treated as a measure of damping and is consistent with the results of fig 9,
i.e. as initial displacement increases the damping becomes large with a
proportionality to the pendulum's velocity. 

The resistive force being
proportional to higher powers of velcoity has been reported earlier
also\cite{r3,r4,r13,r14,r15,r16}. A system is reported to have a
constant friction (${\rm \gamma=0}$) or a linear dependence of velocity 
(${\rm \gamma=1}$) or a quadratic depenence of velocity (${\rm \gamma=2}$).
Corresponding to which the pendulum's amplitude decays linearly,
exponentially and inverse power decay respectively with time. It hence maybe
concluded that for our pendulum sent into motion by initial displacements
${\rm \Theta \geq 45^o}$, the damping force is proportional to ${\rm
v^\gamma}$ where ${\rm \gamma>1}$.

\section{Conclusion}

A simple experiment of setting a suspended bar magnet into oscillations, is
a rich source of information. Not only does it give exposure to Faraday's
induction law and a basic understanding of induced emf's dependence on angle
of oscillation, it enables us to study the damping effects on the pendulum.
This method is better than previously used methods since the measuring
technique does not introduce additional contributions to damping. When the
oscillation imparted to the pendulum is very large, the damping effect is
also strong with the damping force being proportional to ${\rm v^\gamma}$,
where ${\rm \gamma >1}$ and 'v' is the pendulum's velocity. This brings down
the oscillation amplitude of the pendulum and it's velocity. As the velocity
becomes low, the resistive force acting on the pendulum changes it's nature
and becomes proportional to 'v'. Considering the rich information obtained
from the experiment and the simplicity of the experiment, it allows the
method to be easily implemented as a routine experiment in undergraduate
laboratories.


\section*{Acknowledgements}

The authors would like to express their gratitude to the lab technicians of
the Department of Physics and Electronics, SGTB Khalsa College, for the help
rendered in carrying out the experiment.

\newpage
\section*{Appendix}
\begin{table}[h]
\begin{center}  
{\bf Table 1.} Program used to collect data.
\end{center}    
\begin{center}  
\begin{tabular}{||c|c|c||c|c|c||}\hline\hline
Address  & Mnemonics  & Hex Code & Address  & Mnemonics  & Hex Code\\
\hline
C400  & LXI SP	 &  31	 & C40A  & 01H  &  01 \\
C401  & 00H	 &  00	 & C40B  & JZ   &  CA \\
C402  & C3H    	 &  C3	 & C40C  & 07H  &  07 \\
C403  & MVI A	 &  3E   & C40D  & C4H  &  C4 \\
C404  & 00H	 &  00	 & C40E  & IN	&  DB \\
C405  & OUT	 &  D3	 & C40F  & 09H	&  09 \\
C406  & 08H	 &  08	 & C410  & PUSH PSW &  F5 \\
C407  & IN	 &  DB	 & C411 & JMP	& C3 \\
C408  & OBH	 &  OB	 & C412	& 07H	& 07 \\
C409  & ANI	 &  E6	 & C413 & C4H 	& C4 \\
\hline\hline
\end{tabular}
\end{center}
\end{table}  

\begin{figure}[h]
\begin{center}
\epsfig{file=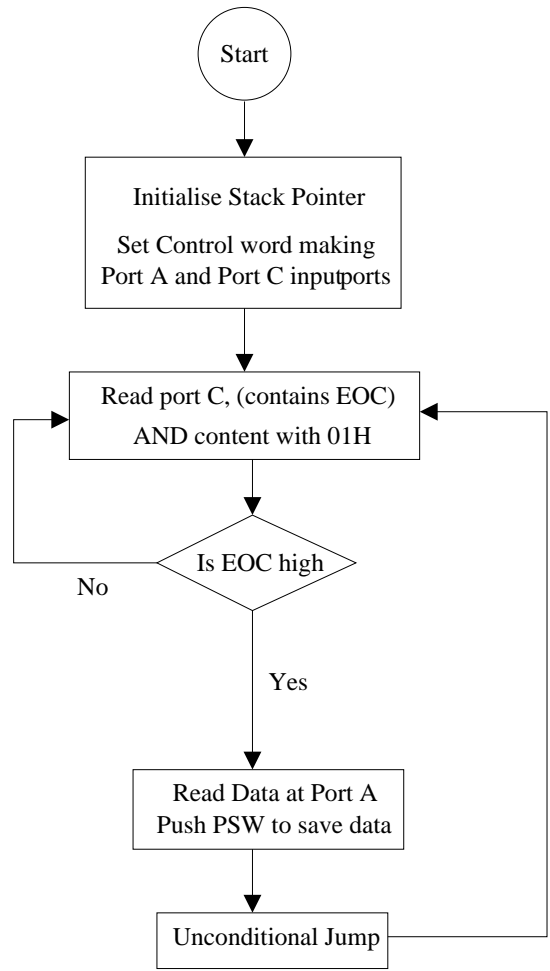,width=3in}
\vskip -.3cm
\caption{Flowchart.}
\vskip -1cm
\end{center}
\label{fig:5xx}
\end{figure}

\end{document}